\documentclass[aps,prb,amsmath,amssymb,twocolumn,superscriptaddress,floatfix]{revtex4-2}
\usepackage{graphicx}
\usepackage{hyperref}
\usepackage{dcolumn}
\usepackage{tabularx}
\usepackage{bm}
\usepackage[version=4]{mhchem}
\usepackage[dvipsnames]{xcolor}
\hypersetup{pdftitle={MnSn(OH)6},pdfauthor={Kaushick},pdfsubject={Frustrated magnetism},pdfdisplaydoctitle}

\def\mshH{MnSn(OH)$_\text{6}$}
\def\mshD{MnSn(OD)$_\text{6}$}

\def\Cp{\ensuremath{c_P}}
\def\Cmag{\ensuremath{c_\text{mag}}}
\def\Smag{\ensuremath{\Delta S_\text{mag}}}

\begin{document}

\title{Disordered ground state in the three-dimensional face-centred frustrated spin-$\frac{5}{2}$ system \mshH}

\author{Kaushick K.\ Parui}
\author{Anton A.\ Kulbakov}
\affiliation{Institut f\"ur Festk\"orper- und Materialphysik, Technische Universit\"at Dresden, 01062 Dresden, Germany}

\author{Ellen H{\"a}u{\ss}ler}
\affiliation{Fakult{\"a}t f{\"u}r Chemie und Lebensmittelchemie, Technische Universit{\"a}t Dresden, 01062 Dresden, Germany}

\author{Nikolai S.\ Pavlovskii}
\author{Aswathi Mannathanath Chakkingal}
\affiliation{Institut f\"ur Festk\"orper- und Materialphysik, Technische Universit\"at Dresden, 01062 Dresden, Germany}

\author{Maxim Avdeev}
\affiliation{Australian Nuclear Science and Technology Organisation, Lucas Heights, NSW 2234, Australia}
\affiliation{School of Chemistry, The University of Sydney, Sydney, NSW 2006, Australia}

\author{Roman Gumeniuk}
\affiliation{Institut für Experimentelle Physik, TU Bergakademie Freiberg, 09596 Freiberg, Germany}

\author{Sergey Granovsky}
\affiliation{Institut f\"ur Festk\"orper- und Materialphysik, Technische Universit\"at Dresden, 01062 Dresden, Germany}

\author{Alexander Mistonov}
\affiliation{Institut f\"ur Festk\"orper- und Materialphysik, Technische Universit\"at Dresden, 01062 Dresden, Germany}

\author{Sergei A.\ Zvyagin}
\affiliation{Dresden High Magnetic Field Laboratory (HLD-EMFL), Helmholtz-Zentrum Dresden-Rossendorf (HZDR), 01328 Dresden, Germany}

\author{Thomas Doert}
\affiliation{Fakult{\"a}t f{\"u}r Chemie und Lebensmittelchemie, Technische Universit{\"a}t Dresden, 01062 Dresden, Germany}
%\affiliation{W\"urzburg-Dresden Cluster of Excellence on Complexity and Topology in Quantum Matter\,---\,ct.qmat, Technische Universit\"at Dresden, 01062 Dresden, Germany}

\author{Dmytro S.\ Inosov}
\email{dmytro.inosov@tu-dresden.de}
\affiliation{Institut f\"ur Festk\"orper- und Materialphysik, Technische Universit\"at Dresden, 01062 Dresden, Germany}
\affiliation{W\"urzburg-Dresden Cluster of Excellence on Complexity and Topology in Quantum Matter\,---\,ct.qmat, Technische Universit\"at Dresden, 01062 Dresden, Germany}

\author{Darren C.\ Peets}
\email{darren.peets@tu-dresden.de}
\affiliation{Institut f\"ur Festk\"orper- und Materialphysik, Technische Universit\"at Dresden, 01062 Dresden, Germany}

\begin{abstract}
Frustrated magnetism in face-centered cubic (fcc) magnetic sublattices remains underexplored but holds considerable potential for exotic magnetic behavior. Here we report on the crystal structure and the magnetic and thermodynamic properties of the $A$-site-vacant hydroxide double perovskite \mshH. Despite dominant antiferromagnetic interactions among $\text{Mn}^{2+}$ moments, evidenced by a negative Curie-Weiss temperature, the lack of a sharp thermodynamic transition down to 350\,mK implies the absence of long-range magnetic order. However, a broad hump in the specific heat at 1.6~K suggests short-range correlations. Neutron diffraction at low temperatures confirms the presence of three-dimensional antiferromagnetic correlations, manifested as diffuse magnetic scattering with a correlation length $\xi = 24.66$\,\AA\ and magnetic propagation vectors \mbox{$\mathbf{k}=(\frac{1}{2}\,\frac{1}{2}\,\frac{1}{2})$} and {$(0~0.625~0)$} at 20~mK. 

\end{abstract}
\maketitle 

\section{Introduction}
Frustrated magnetism has emerged as --- and remains --- a key focus in condensed matter physics, especially after Anderson's prediction of a resonating valence bond (RVB) spin liquid state that suppresses N\'eel order \cite{Anderson1973}. This concept has sparked a surge in both theoretical and experimental research in recent decades. The necessary competition of interactions is perhaps most easily achieved through geometrical frustration, where the crystal structure itself prevents the spins from readily selecting a ground state~\cite{Ramirez1994, Batista2016, Schmidt2017}.  

The face-centered-cubic (fcc) magnetic sublattice with antiferromagnetic (AFM) interactions exemplifies a geometrically frustrated three-dimensional (3D) lattice~\cite{Cook2015}.  This lattice is built up of tetrahedra as in the pyrochlore lattice, but they are edge sharing rather than corner sharing.  A largely unexplored realization of the fcc lattice is the $A$-site-vacant hydroxide double perovskites, $\square_2(BB')(\text{OH})_6$ (where $B$ and $B'$ are magnetic and non-magnetic cations, respectively, and $\square$ denotes the vacant $A$ site). The {\itshape B} and {\itshape B'} sublattices each comprise edge-sharing tetrahedra linked through hydroxide groups, fostering competing AFM interactions that are inherently frustrated.

Theoretical studies of the fcc lattice, although limited, predict a rich magnetic phase diagram driven by frustration and competing interactions. In the framework of the $J_1$-$J_2$ Heisenberg model, commensurate antiferromagnetic ordering vectors $(\frac{1}{2}\,\frac{1}{2}\,\frac{1}{2})$, $(1\,0\,0)$ and $(1\,\frac{1}{2}\,0)$ emerge alongside the possibility of a quantum spin-liquid state~\cite{Revelli2019, Oitmaa2023}. Incorporating third-neighbor interactions ($J_3$) further enriches this landscape, introducing incommensurate spin-spiral phases such as $(q\,0\,0)$, $(q\,q\,q)$ and $(q\,q\,0)$~\cite{Balla2020}. These predictions underscore the potential of fcc-lattice systems---including hydroxide double perovskites---to realize exotic magnetic ground states.

Mn atoms on an fcc lattice have a well-documented tendency to form complex AFM structures due to frustrated exchange interactions. Classical examples such as MnO, MnSe and $\alpha$-MnS exhibit type-II order~\cite{Shull1951, Corliss1956} (also known as ``checkerboard'' ordering), where \ce{Mn^2+} spins align ferromagnetically along $\langle100\rangle$ directions, with adjacent planes antiferromagnetically aligned, driven by dominant nearest-neighbor Mn--$X$--Mn superexchange ($J_1$). In contrast, dichalcogenides show varied behavior: \ce{MnS_2} exhibits type-III order~\cite{Hastings1959,Chatterji2019}, which retains an in-plane AFM arrangement among nearest neighbors but also features antiferromagnetic stacking of next-nearest-neighbor spins along the out-of-plane $[001]$ direction. Meanwhile, \ce{MnSe_2} and \ce{MnTe_2} adopt type-I (``rock-salt'') order~\cite{Hastings1959}, with ferromagnetic $\langle111\rangle$ planes stacked antiferromagnetically. These differences reflect the growing influence of ferromagnetic second-nearest-neighbor exchange ($J_2$) in this series, underscoring the critical role of indirect superexchange pathways in shaping the magnetic ground states of fcc \ce{Mn^2+} systems.

Hydroxide double perovskites also offer a unique platform to study complex magnetic interactions and the expected tunability of magnetic properties by varying the magnetic cations, similar to that found in conventional transition-metal double perovskites~\cite{Welch2025}. For instance, combining spin-only $3d$ transition-metal ions (with quenched orbital moments) with strongly spin-orbit coupled $4d$ or $5d$ ions creates spin-orbit--entangled quantum materials~\cite{Jin2022}. The resulting interplay of strong correlations, geometrical frustration, spin-orbit coupling, and quantum fluctuations is known to generate a rich spectrum of magnetic phases in double perovskites, including quantum spin liquids (QSL), spin ices, and spin glasses \cite{Morrow2013, Morrow2015, Gangopadhyay2016, Morrow2016, Manna2016, Terzic2017, Bhowal2018, Morrow2018, Aczel2019, Revelli2019, Jin2022,Kulbakov2025b}.

Moreover, in these systems, superexchange interactions between $3d$ and $5d$ ions deviate from traditional Goodenough–Kanamori rules~\cite{Morrow2013}, and chemical pressure can be used to tune bond angles, potentially inducing a transition from antiferromagnetic to ferromagnetic behavior~\cite{Morrow2015}.  Mn-based analogues exhibit diverse ordering phenomena---such as a fourfold degenerate non-collinear antiferromagnetic order in Ba$_{2}$MnMoO$_{6}$~\cite{Coomer2023}, and type-I and II order in Ba$_{2}$MnTeO$_{6}$~\cite{Mustonen2020} and Ba$_{2}$MnWO$_{6}$~\cite{Mutch2020}, respectively. The choice between type-I and type-II order is governed by the relative strengths of first‑ and second‑neighbor exchange on the fcc lattice ($J_1$ vs.\ $J_2$), as well as single‑ion anisotropy; bond angles, which can be tuned by chemical pressure; and $B$--O--$B'$ bond lengths. Short-range magnetic correlations persisting well above the ordering temperature in these compounds further underscore the role of geometric frustration and strong magnetic interactions.
%Spin-orbit coupling induces spin anisotropy, which partially alleviates magnetic frustration and provides a means to control ground-state properties. Additionally, spin-orbit coupling favors non-Heisenberg (Kitaev-type) bond-directional exchange \cite{Trebst2022}, offering a pathway to spin-liquid states via exchange frustration on the fcc lattice~\cite{Revelli2019}. The phase diagram for the fcc lattice, calculated by Revelli \textit{et al.}~\cite{Revelli2019}, identifies a stable quantum spin-liquid phase, with the double-perovskite iridate Ba$_{2}$CeIrO$_{6}$ being in close proximity to this phase. 
These insights suggest that the unique magnetic behavior observed in double perovskites may extend to hydroxide perovskites, opening new avenues for exploring exotic quantum magnetic states.

\mshH\ was discovered as a natural mineral in two distinct crystallographic forms: cubic wickmanite~\cite{Moore1967}, with a $Pn\overline{3}$ space group and an $a^{+}a^{+}a^{+}$ tilt scheme, and tetragonal tetrawickmanite~\cite{White1973}, with a $P4_2/n$ space group and an $a^{+}a^{+}c^{-}$ tilt system. Here, $a^{+}a^{+}a^{+}$ indicates equal in-phase rotations of the octahedra about all three crystallographic axes, while $a^{+}a^{+}c^{-}$ represents in-phase tilts along the $a$ and $b$ axes and a reverse tilt along the $c$ axis. Crystal structure studies of synthetic polycrystalline wickmanite revealed a single type of cavity, where hydrogen atoms are disordered across two positions~\cite{Basciano1998}. This arrangement leads to ``ice rules'' and correlated structural disorder\,---\,a common trait of hydroxide perovskites~\cite{Kulbakov2025a}. In contrast, tetrawickmanite, which shares a similar topology, exhibits proton disorder across four positions, with one ordered hydrogen forming two distinct cavity environments\,---\,one akin to wickmanite’s four-membered hydrogen-bonding ring and the other featuring distorted crankshaft-type motifs~\cite{Lafuente2015}. The impact of correlated hydrogen disorder on magnetism in this family remains uncertain, but disorder is known to strongly influence magnetic correlations. In magnetic charge ices, for example, geometric frustration combined with correlated bond disorder can lead to emergent phenomena such as spin nematic order in pyrochlores~\cite{Hemmatzade2024}. Previous low-temperature magnetic measurements on synthetic polycrystalline wickmanite found that it remained paramagnetic down to 2\,K~\cite{Neilson2011}.

In this study, we report the synthesis and characterization of tetragonal \mshH\ (tetrawickmanite). Our bulk magnetic measurements, specific heat data, and diffuse magnetic scattering studies indicate the absence of magnetic long-range order (LRO), with the emergence of short-range magnetic correlations around 1.6\,K. Analysis of the diffuse magnetic scattering using the reverse Monte Carlo (RMC) method reveals 3D antiferromagnetic short-range correlations extending down to 20\,mK. Hence, \mshH\ may offer a useful platform for investigating the influence of correlated hydrogen disorder on the magnetic properties.  

\section{Experimental Methods}

Polycrystalline samples of \mshH\ were prepared via a hydrothermal route under autogenous pressure. Initially, a stoichiometric mixture comprising \mbox{Na$_2$SnO$_3\cdot 3$H$_2$O} (ThermoFisher GmbH, 96\%) and MnCl$_2\cdot 4$H$_2$O (ThermoFisher GmbH, 99.997\%) in a 1:1 molar ratio was thoroughly mixed in an agate mortar. The resulting homogeneous mixture was subsequently transferred to a 50-mL Teflon-lined stainless-steel autoclave, to which 20\,mL of deionized water were added. The autoclave was sealed and held at 50\,$^\circ$C in a convection drying oven for 5 days. Following the heating phase, the autoclave was allowed to cool to room temperature naturally.  The resulting product was filtered, then washed thoroughly with deionized water to remove soluble impurities such as NaCl. Finally, the washed powder was dried in a vacuum furnace at room temperature. A deuterated batch was prepared for neutron scattering experiments using D$_2$O (Acros Organics, 99.8\,at.\,\% D) in place of H$_2$O, aiming to reduce the incoherent scattering from hydrogen.

A powder x-ray diffraction (XRD) pattern was recorded on \mshD\ at room temperature using a STOE Stadi P diffractometer in transmission mode. The instrument utilized Ag-K$\alpha_1$ radiation ($\lambda = 0.559$\,\AA) to scan an angular range from 4.0$^\circ$ to 73.3$^\circ$ in $2\theta$.

The surface morphology of \mshH\, powder was imaged using an in-lens detector on a Carl Zeiss AG\ Ultra 55 field-emission scanning electron microscope (SEM) at ambient temperature. The powder was sprinkled on a carbon-adhesive tape affixed to a sample puck. Energy-dispersive x-ray spectroscopy (EDX) was conducted with a Bruker\ Quantax EDS system, and data analysis was performed using Bruker's \textsc{Esprit} software package.

Fourier-transform infrared spectroscopy (FTIR) was measured using a Bruker Vertex 70 in attenuated total reflectance (ATR) construction between 400 and 4000\,cm$^{-1}$ with 2\,cm$^{-1}$ resolution on both \mshH\ and \mshD\ powders.

Neutron powder diffraction (NPD) data on \mshD\ were collected over a $2\theta$ angular range from 4.0$^\circ$ to 163.9$^\circ$ at the Echidna diffractometer~\cite{Echidna} located at the OPAL research reactor, Australian Science and Technology Organization (ANSTO), Lucas Heights, Australia. Measurements were performed on a large powder sample comprising several syntheses using 2.44-\AA\ neutrons at temperatures of 4\,K, 500\,mK, and 20\,mK, with collection times of 50, 8 and 50 hours, respectively, and at room temperature using 1.30-\AA\ neutrons. Both the XRD and NPD patterns were analyzed by the Rietveld refinement method~\cite{Rietveld1969} using the  \textsc{FullProf} software package~\cite{FullProf}. The magnetic diffuse neutron scattering pattern was modeled using the Reverse Monte Carlo (RMC) method with the  \textsc{Spinvert} program~\cite{SPINVERT}. The crystal structure was visualized using {\sc Vesta}~\cite{VESTA}.

Temperature-dependent DC magnetization measurements were performed on \mshH\ using a vibrating sample magnetometer (VSM) in a Cryogenic Ltd.\ Cryogen-Free Measurement System (CFMS). The sample, held in a gelatin capsule mounted in a plastic straw, was measured under zero-field-cooled-warming, field-cooled-cooling, and field-cooled-warming conditions. Isothermal magnetization was recorded at 10\,K across a field range of $\pm 14\,\text{T}$. High-field magnetization measurements up to 60\,T at 1.4\,K were conducted at the Hochfeld-Magnetlabor Dresden (HLD), Helmholtz-Zentrum Dresden-Rossendorf (HZDR), in Dresden, Germany, using a pulsed magnet with a rise time of 7\,ms and a total pulse duration of 25\,ms. The magnetization was obtained by integrating the voltage induced in a compensated coil system surrounding the sample\,\cite{Skourski2011}. The ac susceptibility was measured using an Oxford Instruments MagLab System2000, with temperature ranging from 1.5 to 6~K. 

Low-temperature specific-heat measurements were performed on a pressed pellet of \mshH\ using the two-tau relaxation method in a Quantum Design Physical Property Measurement System (PPMS) DynaCool-12 system equipped with a $^3$He refrigerator. Addenda measurements were conducted beforehand to account for contributions from the sample holder and Apiezon N grease.
  
\section{Crystal Structure}

\begin{figure}[t]
  \includegraphics[width=\columnwidth]{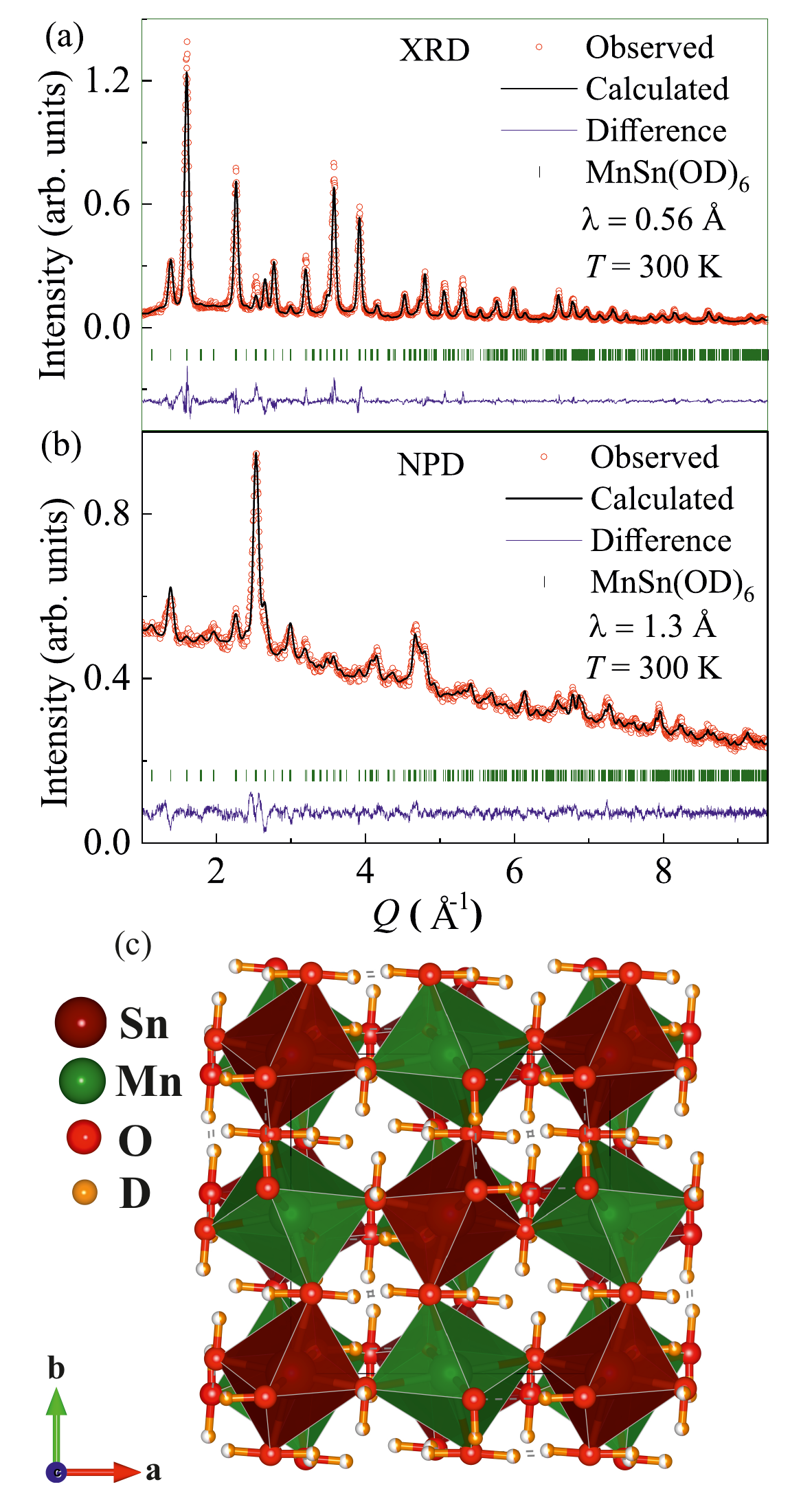}\vspace{-10pt}
  \caption{\label{NPD}Rietveld-refined powder (a) x-ray  and (b) neutron diffraction patterns for \mshD\ at room temperature. (c) Refined crystal structure in tetragonal $P4_2/n$ from Echidna NPD data.}
\end{figure}

\begin{figure}[t]
  \includegraphics[width=\columnwidth]{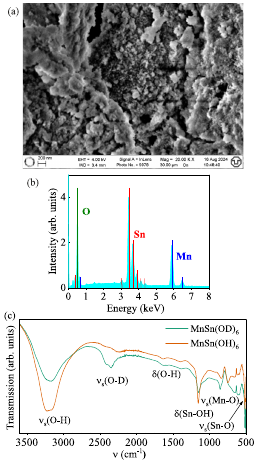}\vspace{-10pt}
  \caption{\label{SEM_EDX_IR}(a) High-resolution SEM image of our \mshH\ powder. (b) EDX spectrum, illustrating the elemental composition of \mshH. (c) FTIR spectra of  \mshH\, and \mshD\ at room temperature.}
\end{figure}

The crystal structure of \mshD\ was investigated using x-ray and neutron diffraction at room temperature, as shown in Fig.~\ref{NPD}. Rietveld refinement confirms the compound is single-phase, crystallizing in tetragonal symmetry with space group $P4_2/n$ (\#~86), consistent with earlier x-ray observations on a single-crystal sample of tetrawickmanite \cite{Lafuente2015}. The refined structural parameters and goodness-of-fit factors are summarized in Tables \ref{Summary}, \ref{XRD_Stoe}, and \ref{NPD_Echidna} in Appendix~\ref{appA}, and crystallographic information files (CIFs) are available in 
the ancillary files online, see Appendix~\ref{supp}. 
%the Supplemental Material online\,\cite{SuppMnSn}.
The refined crystal structure of \mshD, illustrated in Fig.~\ref{NPD}(c), reveals a 3D network characterized by alternating corner-sharing $[\text{Mn}^{2+}(\mathrm{OD})_6]$ and $[\text{Sn}^{4+}(\mathrm{OD})_6]$ octahedra. The structure features three distinct oxygen atoms, each bonded to five unique deuterium positions. Interestingly, only the O3--D5 bond exhibits order, while the other O--D bonds display hydrogen disorder. This arrangement of hydrogen atoms follows ``ice rules'' analogous to those observed in hexagonal water ice structures\,\cite{Bernal1933,Pauling1935}. 

Surface morphology and elemental composition were analyzed using SEM and EDX. The SEM image in Fig.~\ref{SEM_EDX_IR}(a) depicts nanometer-sized particles in the polycrystalline sample. The EDX spectrum shown in Fig.~\ref{SEM_EDX_IR}(b) identifies Mn, Sn and O, with no detectable impurities, and shows a Mn:Sn ratio of 1:1.12, near the theoretical 1:1 ratio. Diffraction did not find any cation nonstoichiometry, so this is attributed to the uncertainty of the EDX technique and the difficulties of working with a fine, electrically insulating powder.

\begin{figure*}
  \includegraphics[width=\textwidth]{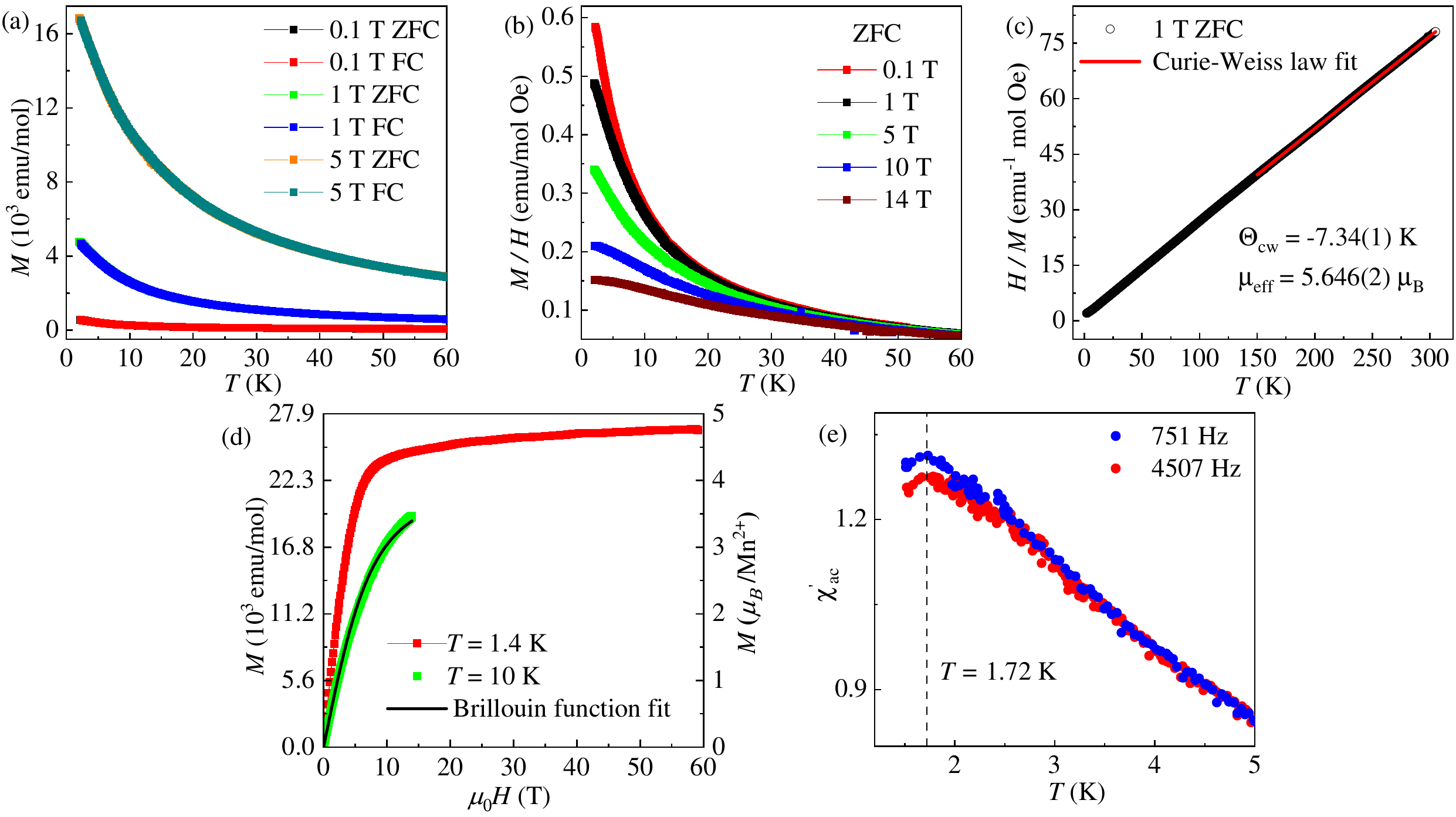}
  \caption{\label{Magnetization}(a) Temperature dependence of \mshH\ magnetization under ZFC and FC conditions at selected fields. (b) ZFC $M/H$ as a function of temperature at various fields. (c) Inverse ($H/M$) of the ZFC curve at $\mu_{\text{0}}H = 1$\,T with the solid red line indicating the Curie-Weiss fit. (d) Isothermal magnetization at 1.4 and 10\,K, with the solid black line indicating the fit using the Brillouin function. (e) Temperature dependence of the real part of the ac susceptibility for \mshH\ measured at 751 and 4507\,Hz under zero dc magnetic field.}
\end{figure*}
  
FTIR spectroscopy was performed to clarify details of the bonding. The FTIR spectra of \mshH\ and \mshD\ in Fig.~\ref{SEM_EDX_IR}(c) highlight a key difference: the broad O--D stretching band around 2340\,$\text{cm}^{-1}$ in \mshD, confirming proton replacement by deuterium. The O--H stretching and bending modes are centered around 3200 and 1637\,$\text{cm}^{-1}$, respectively. The Sn--OH bending mode appears at 1150\,$\text{cm}^{-1}$, and metal-oxygen stretching modes are seen as multiple absorption bands in the 400--700\,$\text{cm}^{-1}$ range. The FTIR spectrum acts as a fingerprint of the material, and subtle changes may help distinguish the polymorphs of \mshH.

\section{Magnetization \& ac susceptibility}

Magnetization measurements were performed as a function of temperature and magnetic field to probe exchange interactions and spin dynamics in \mshH. The temperature dependence of the zero-field-cooled (ZFC) and field-cooled (FC) dc magnetization ($M$) of \mshH\ in an applied magnetic field ($H$) of 1\,T is shown in Fig.~\ref{Magnetization}(a), while Fig.~\ref{Magnetization}(b) presents the temperature-dependent ZFC $M/H$ in various magnetic fields. No divergence of the ZFC from the FC data is observed down to 2\,K in any field, indicating the absence of spin freezing. $M/H$ increases smoothly on cooling and is suppressed by increasing magnetic field $H$. Furthermore, \mshH\, shows no signs of long-range magnetic order up to $14\,\text{T}$ and down to 2\,K, behaving as a paramagnet and confirming the absence of any magnetic phase transition down to 2\,K. The data from 150--300\,K measured in a 1-T applied field under ZFC conditions is fitted using the Curie-Weiss law given by
\setlength{\abovedisplayskip}{5pt}
\setlength{\belowdisplayskip}{5pt}
\begin{align}
\chi = \chi_0 + \frac{C}{(T - \theta_{\text{CW}})}
\end{align}
where $\chi_0$ is a temperature-independent susceptibility term that includes contributions from both diamagnetism and van~Vleck paramagnetism, and $C$ and $\theta_{\text{CW}}$ are the Curie constant and Weiss temperature, respectively. From Fig.~\ref{Magnetization}(c), it is quite evident that \mshH\ exhibits Curie-Weiss behavior. The best fit results in $\chi_0 = 4.754\times10^{-5}$\,emu/mol$_\text{Mn}$, $\theta_{\text{CW}} = -7.34(1)$\,K, and $C = 3.98(4)$\,emu\,K/mol$_\text{Mn}$\,Oe. The small, negative $\theta_{\text{CW}}$ suggests weak net AFM interactions between $\text{Mn}^{2+}$ spins. The effective paramagnetic moment obtained experimentally is $\mu_{\text{eff}} = 5.646(2)\,\mu_{\text{B}}/\text{Mn}^{2+}$, which is in good agreement with the theoretical spin-only $S=5/2$ value of $5.92\,\mu_{\text{B}}/\text{Mn}^{2+}$ for free $\text{Mn}^{2+}$ ions [$\mu_{\text{eff}} = g_J \sqrt{J (J + 1)}\, \mu_{\text{B}}$].

\begin{figure*}
  \includegraphics[height=0.45\textheight]{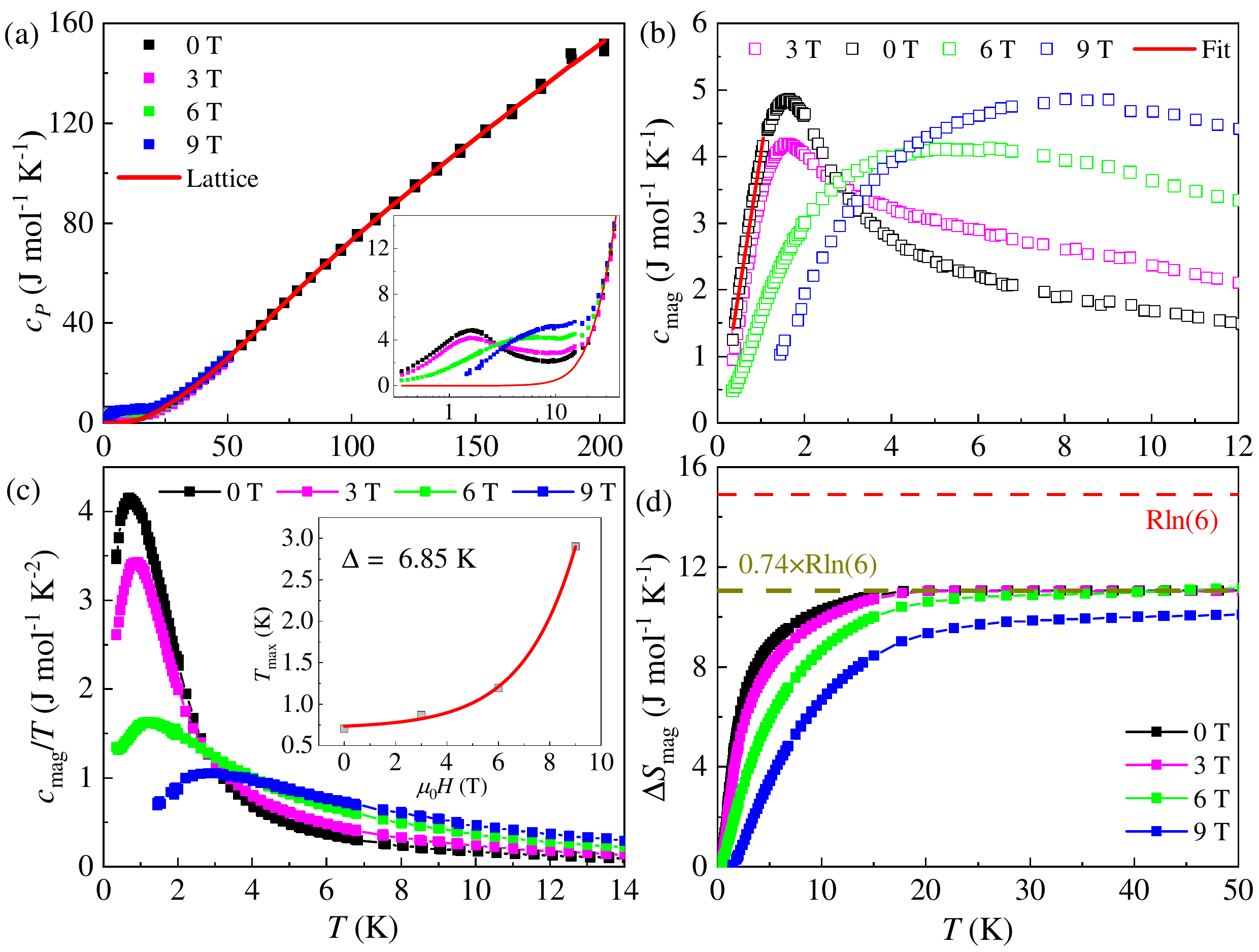}
  \caption{\label{Specific_heat}(a) Temperature dependence of total specific heat (\Cp) of \mshH\ measured down to 350\,mK under various magnetic fields. The solid line represents the lattice contribution, fitted using the Debye-Einstein model. Inset: A zoomed-in view of the low-$T$ \Cp\ anomaly. (b) Magnetic specific heat (\Cmag) as a function of temperature in several magnetic fields, with solid line indicating power-law fit as described in the text. (c) Temperature dependence of $\Cmag/T$ under different magnetic fields. Inset: The shift in the broad maximum with increasing magnetic field, fitted using Eq.~\ref{Tmax}. (d) Magnetic entropy (\Smag\ as a function of temperature, with dotted lines representing both theoretical expectations and experimental values of entropy.}
\end{figure*}

Figure~\ref{Magnetization}(d) shows the isothermal magnetization curves at 1.4 and 10\,K. Below 10\,T, $M$ increases linearly, while above 10\,T, it becomes non-linear and tends to saturate, indicating weak exchange interactions in \mshH. The saturation magnetization, estimated by \mbox{$M_{\text{s}} = g_JJ\mu_{\text{B}}$} for $\text{Mn}^{2+}$ ions ($M_{\text{s}} = 5\,\mu_{\text{B}}/$\text{Mn}$^{2+}$), reaches a maximum of  4.76\,$\mu_{\text{B}}/$\text{Mn}$^{2+}$ at 1.4\,K, which is close to the expected value. The slight shortfall in magnetization may stem from weak AFM spin arrangements, or from calibration issues with the CFMS. The isothermal magnetization curve in the paramagnetic state at 10\,K is fitted using the relation ${M} = {M_s}B_J(y)$, where the Brillouin function $B_J(y)$ is defined as
\setlength{\abovedisplayskip}{5pt}
\setlength{\belowdisplayskip}{5pt}
\begin{align}
B_J(y) = \left[\frac{2J+1}{2J}\coth\left(\frac{y(2J+1)}{2J}\right) - \frac{1}{2J}\coth\left(\frac{y}{2J}\right)\right].
\end{align}
In this context, $y = {g\mu_BJ\mu_0H}/{k_BT}$, with $g$ representing the Land{\'e} $g$ factor. For the fit, $J$ was fixed at $5/2$, leaving $g$ as the sole adjustable parameter. The solid line in Fig.~\ref{Magnetization}(d) represents the Brillouin function fit, yielding a $g$ factor of 1.55, which is significantly lower than the expected value of 2. Furthermore, the fit deviates from the experimental data, indicating that simple Brillouin behavior does not fully capture the system's magnetic response. This discrepancy likely arises from the presence of antiferromagnetic correlations, which are not accounted for in the non-interacting spin model. Similar deviations are also seen in the 1.4-K data. Aside from saturation, no field-induced transitions are observed in \mshH.

\begin{figure*}
  \includegraphics[width=\textwidth]{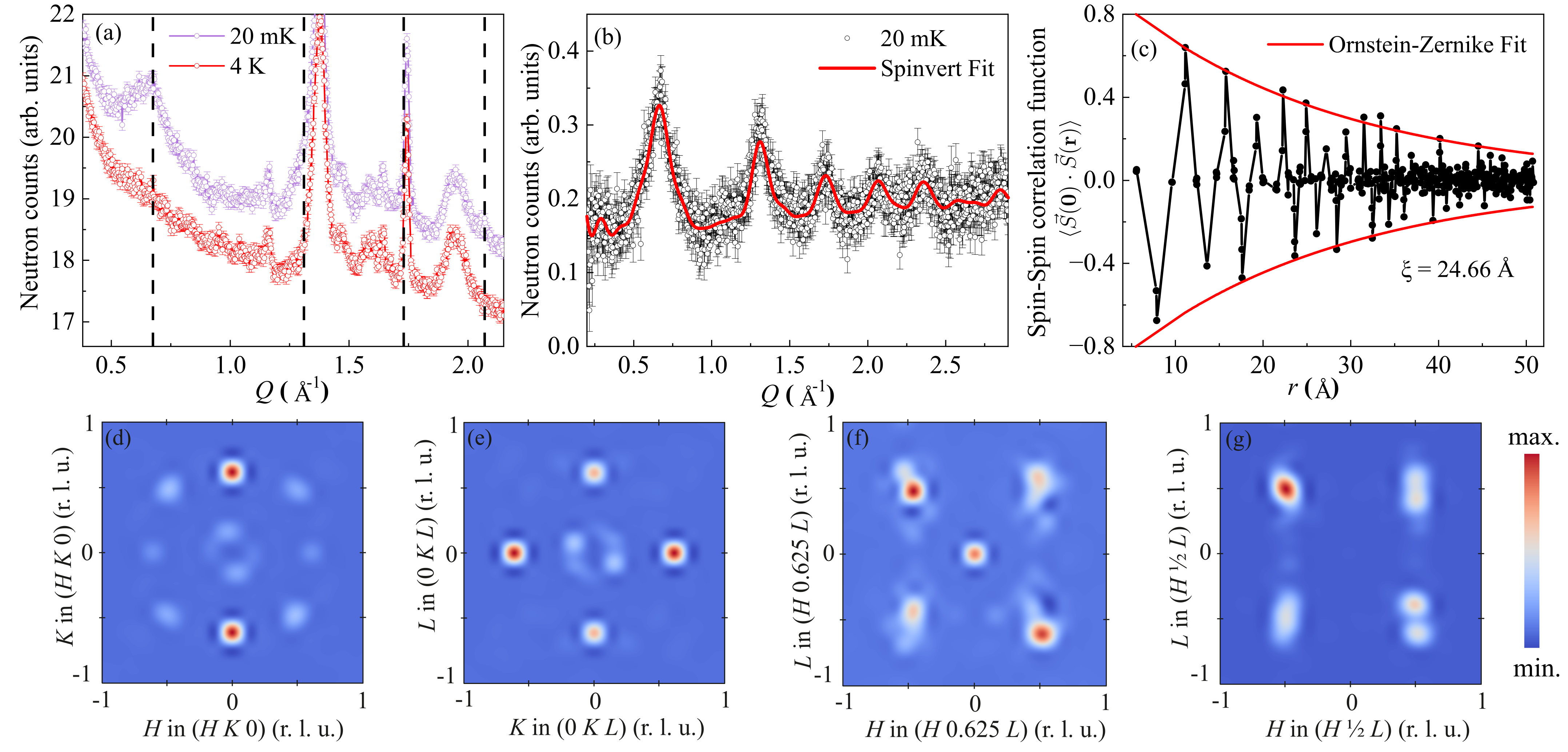}
  \caption{\label{Short_range}(a) Neutron diffraction patterns measured at 20\,mK and 4\,K. The dashed lines mark the positions of broad diffuse magnetic peaks that emerge at low temperatures. (b) \textsc{Spinvert} fits to the diffuse magnetic scattering at 20\,mK. (c) Radial spin-pair correlations based on the \textsc{Spinvert} fit in (b) modeled using the Ornstein-Zernike equation. (d-g)~Reconstructed diffuse magnetic scattering patterns for $(H\,K\,0)$, $(0\,K\,L)$, $(H\,0.625\,L)$ and $(H\,\frac{1}{2}\,L)$ planes, respectively.}
\end{figure*}

A broad hump in the real part of ac susceptibility $\chi'_{\text{ac}}$ is observed at 1.72\,K (Fig.~\ref{Magnetization}(e)). Supported by corresponding features in the specific heat data (see next section), this anomaly is attributed to the onset of short-range magnetic correlations. Notably, the hump position remains unchanged across measured frequencies up to 4507\,Hz, suggesting an absence of glassy behavior in the system. No features were resolved in the imaginary part of the susceptibility $\chi''_{\text{ac}}$.
  
\section{Specific Heat}

Specific heat (\Cp) measurements were performed over a wide range of temperatures and magnetic fields to explore magnetic correlations and low-energy excitations in \mshH. As shown in Fig.~\ref{Specific_heat}(a), no sharp anomaly indicative of a transition to magnetic long-range order was observed down to 350\,mK. Instead, a broad hump centered below 2\,K, illustrated in the inset of Fig.~\ref{Specific_heat}(a), suggests short-range magnetic order, consistent with the hump in ac susceptibility measurements.

In the absence of an isostructural nonmagnetic counterpart, the magnetic contribution to the specific heat (\Cmag) was isolated phenomenologically by subtracting an approximate lattice contribution from the total \Cp. The lattice contribution is modeled using the Debye-Einstein approximation, which includes one Debye and three Einstein terms over the 15–200\,K range, where phonon effects dominate. The model is expressed as
\begin{align}
c_{\text{lattice}}(T) &= f_D c_D(\theta_D, T) + \sum_{i=1}^{3} g_i c_{E_i}(\theta_{E_i}, T).
\label{clattice}
\end{align}

The first term in Eq.~\eqref{clattice} represents the Debye model, accounting for acoustic modes:
\begin{align}
c_D(\theta_D, T)~=~&9NR \left( \frac{T}{\theta_D} \right)^3\int_0^{\theta_D/T}\frac{x^4 e^x}{(e^x - 1)^2} \, dx.% \notag \\
\end{align}
Here, $\theta_D$ represents the Debye temperature, $N$ is the total number of atoms in the formula unit, $R$ is the universal gas constant, and $x$ is defined as $\frac{\hbar \omega}{k_B T}$. The second term in Eq.~\eqref{clattice}, known as the Einstein term, accounts for the optical modes:
\begin{align}
c_{E}(\theta_{E}, T)~=~&3NR \left( \frac{\theta_E}{T} \right)^2 \frac{e^{\theta_E/T}}{(e^{\theta_E/T} - 1)^2}
\end{align}
where $\theta_E$ is the Einstein temperature. The coefficients $f_D$, $g_1$, $g_2$ and $g_3$ are weight factors based on $N$, chosen such that their sum is unity to satisfy the Dulong-Petit law at high temperatures.

As seen in Fig.~\ref{Specific_heat}(a), the estimated lattice contribution fits the \Cp\ data well yielding $f_D=0.1$, $\theta_D=176$\,K, $g_1=0.25$, $g_2=0.2$, $g_3=0.45$, $\theta_{E_1}=324$\,K, $\theta_{E_2}=670$\,K and $\theta_{E_3}=1127$\,K.
Figure~\ref{Specific_heat}(b) displays the resulting magnetic contribution \Cmag\ as a function of temperature, showing a broad hump. Magnetic field suppresses the peak in \Cmag\ and broadens it to higher temperatures. For temperatures below the hump, fitting the data using the power-law relation \mbox{$c_\text{mag} = \alpha T^{\beta}$} yields $\alpha = 4.04\pm0.02$ and $\beta = 1.01\pm0.01$, indicating a nearly linear dependence on $T$ at low temperatures in zero field.
Figure~\ref{Specific_heat}(c) shows the variation of $\Cmag/T$ under different magnetic fields, where the anomaly broadens and is progressively suppressed, yet shifts to higher temperatures with increasing field. This counterintuitive behavior suggests that the field modifies short-range AFM correlations or partially resolves frustration rather than simply suppressing magnetic order. The evolution of the hump temperature $T_\text{max}$ with field, shown in the inset of Fig.~\ref{Specific_heat}(c), is well captured by the empirical relation
\begin{align}
T_{\text{max}}(\mu_0 H) = A \exp\left(\frac{\mu_0 H}{\mu_0 H_0}\right) + T_0
\label{Tmax}
\end{align}
with fit parameters $A = 0.027(1)$\,K, $T_0 = 0.71(6)$\,K and $\mu_0H_0 = 2.04(22)$\,T. The associated energy scale can be estimated using the Zeeman relation \mbox{$\Delta = \frac{g\mu_B \mu_0 H_0  S}{k_B}$ which yields $\Delta = 6.85$\,K} (for $g=2$, $S=\frac{5}{2}$). This energy scale closely matches the magnitude of the Curie-Weiss temperature $|\theta_{\text{CW}}| = 7.34(1)$\,K, suggesting that the characteristic field $\mu_0 H_0$ corresponds to the energy required to overcome antiferromagnetic correlations among localized spins. This reinforces the interpretation of $T_\text{max}$ as a signature of field-driven perturbation of short-range correlated AFM spins.

Magnetic entropy (\Smag), calculated as \mbox{$\Smag = \int_{T'}^T \frac{\Cmag}{T} \text{d}T$} (where $T'$ is the lowest measured temperature), is depicted in Fig.~\ref{Specific_heat}(d) as a function of temperature. It saturates around 20\,K at zero field, reaching approximately 74\% of the expected entropy value $R \ln(6) = 14.89\,\text{J/mol} \cdot \text{K}$ for \(S = \frac{5}{2}\) spins. This observed reduction in entropy is also noted in other frustrated systems \cite{Bag2021,Bhattacharya2024,Singh2024,Aswathi2025} and may result from short-range magnetic correlations, overestimation of the lattice contribution, or the release of entropy at lower temperatures not covered by our measurements. This implies potential magnetic correlations at ultra-low temperatures among $\text{Mn}^{2+}$ ions in \mshH.

\section{Short-range Magnetic Correlations}

To clarify the presence of short-range magnetic spin-spin correlations, we conducted neutron diffraction experiments at low temperatures --- see Fig.~\ref{Short_range}(a). Below 4\,K, broad diffuse magnetic peaks emerge with maxima at $Q \sim 0.66$, $1.31$, $1.73$, $2.07$ and $2.35\,\text{\AA}^{-1}$, as seen in Fig.~\ref{Short_range}(b) where we have subtracted the 4-K data as a background to isolate the magnetic scattering at 20\,mK. While these form clear peaks, they are significantly broader than the structural Bragg peaks, with estimated widths (FWHM) significantly exceeding the instrumental resolution. For example, at $Q \sim 0.66\,\text{\AA}^{-1}$, the FWHM of the diffuse peak is 4.43$^\circ$ compared to 0.45$^\circ$ for Echidna. These peaks gain intensity on cooling down to 20\,mK. The absence of sharp magnetic Bragg peaks down to 20\,mK confirms the lack of long-range magnetic ordering in \mshD. Similar diffuse magnetic features have been observed in other double perovskites \cite{Mustonen2024}.

The shape of the diffuse peak reflects the dimensionality of the magnetic order. Three-dimensional magnetic correlations produce a symmetric peak shape, often described by the Lorentzian function. In the case of \mshD, we indeed observe a symmetric peak shape, indicating 3D short-range magnetic correlations.   

To gain deeper insights into the microscopic spin-spin correlations in \mshD, we employed the RMC method using the \textsc{Spinvert} program \cite{SPINVERT}. While a simple Lorentzian fit provides basic correlation lengths, \textsc{Spinvert} offers a more comprehensive analysis. This program fits magnetic diffuse scattering powder data from a randomly initialized spin configuration, without prior knowledge of the spin Hamiltonian. This method has been effectively used in various frustrated magnetic systems \cite{Paddison2012, Kulbakov2022b, Mustonen2024, Saha2023}. It extracts radial spin-spin correlation functions in real space using \textsc{Spincorrel} and generates diffuse magnetic scattering profiles with the \textsc{Scatty} package\,\cite{Scatty}.

\begin{figure}
  \includegraphics[width=\columnwidth]{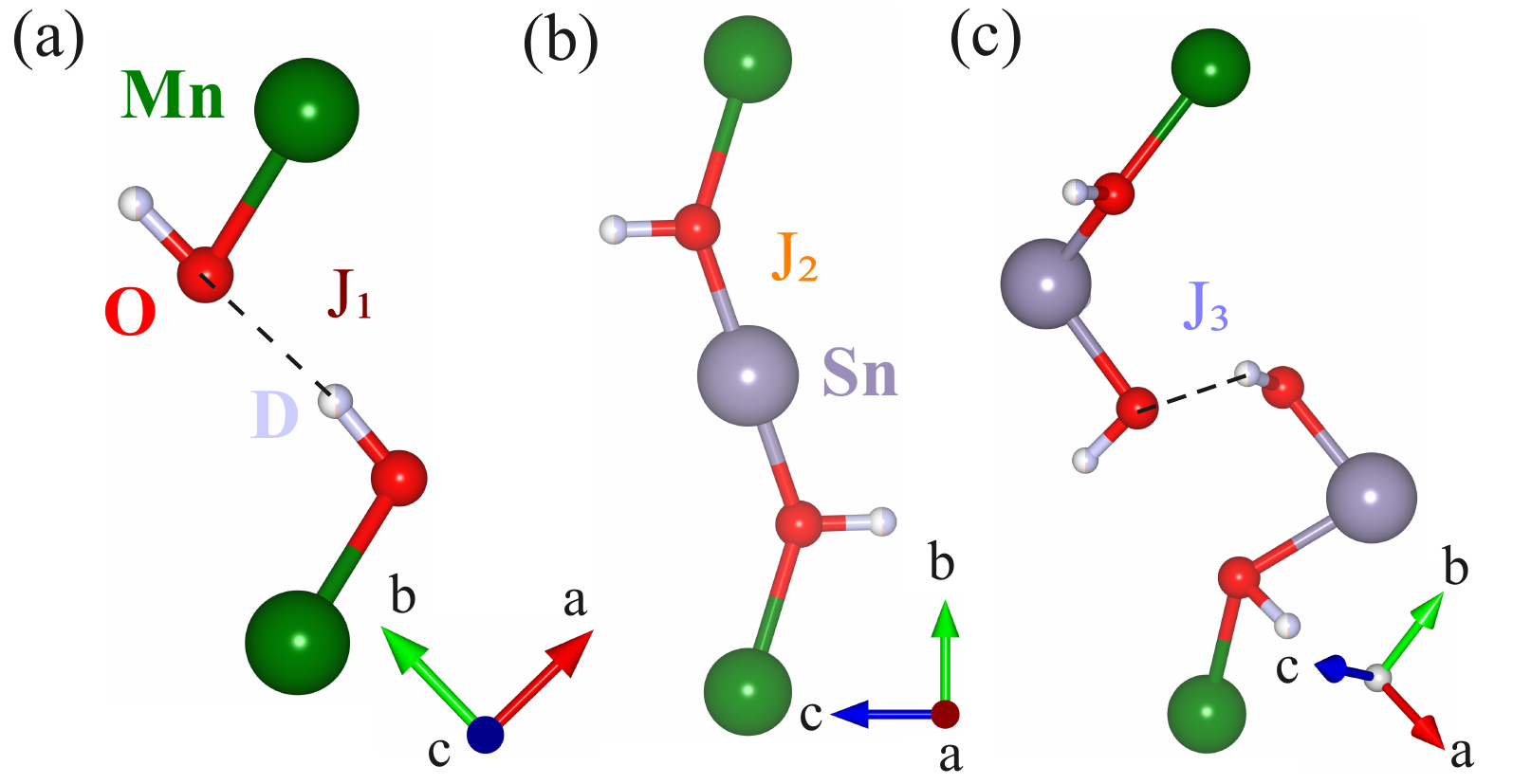}%\vspace{-10pt}
  \caption{\label{Magnetic_Pathway}(a-c) Representative structural configurations corresponding to $J_1$,  $J_2$, and $J_3$ exchange pathways, respectively, as observed in \mshD.}
\end{figure}

In our analysis, we used a supercell of $13 \times 13 \times 13$ unit cells, comprising 8788 spins. During refinement, spin orientations were adjusted (1000 times per spin) while the spin positions remained fixed, to reproduce the experimental data. This procedure was conducted 15 times to confirm the reliability of the results. 

The magnetic diffuse scattering at 20\,mK is overlaid with the fitted curve in Fig.~\ref{Short_range}(b). From this fit, the spin-pair correlations for \mshD, averaged over 15 independent calculations, are presented in Fig.~\ref{Short_range}(c). The spin-pair correlation function \mbox{$\langle \vec{S}(\mathbf0) \cdot \vec{S}(\mathbf r) \rangle$} quantifies the correlation between two $\text{Mn}^{2+}$ spins as a function of lattice distance $r$. The magnitude of the spin-pair correlation function indicates the degree of spin alignment. A magnitude of zero suggests that the magnetic ions are uncorrelated, orthogonal, or sum to zero for that distance, while a magnitude of $\pm 1$ signifies strong correlation, with ($+1$) denoting parallel (FM) and ($-1$) representing antiparallel (AFM) alignment of spins. Each data point in Fig.~\ref{Short_range}(c) represents a Mn--Mn distance, revealing an oscillation between AFM and FM alignment as a function of distance: the shortest Mn--Mn distance at $r_\text{avg} \approx 5.55$ and 5.57\,\AA\ shows zero net correlation ($\langle \mathbf{S}_0 \cdot \mathbf{S}_1 \rangle = 0$), while strong AFM correlations are observed at $r_\text{avg} \approx 7.82$ and 7.88\,\AA. At $r_\text{avg} \approx 9.6$ and 9.63\,\AA, the correlations are again zero, then at $r_\text{avg} \approx 11.10$ and 11.14\,\AA, they become strongly FM. This alternating pattern suggests correlations with a strong AFM component, and they are seen to decay with increasing distance $r$, ultimately vanishing at $\sim\!40$~\AA.  

Using these spin configurations, the magnetic scattering intensities for the $(H\,K\,0)$, $(0\,K\,L)$, $(H\,0.625\,L)$ and $(H\,\frac{1}{2}\,L)$ planes in a single crystal were simulated, as shown in Figs.~\ref{Short_range}(d-g). All scattering planes exhibit symmetric scattering, indicating isotropic magnetic correlations and 3D magnetic interactions, as expected from the nearly cubic crystal structure. The maxima in the reciprocal space planes align with the propagation vectors \mbox{$\mathbf{k}_1=(\frac{1}{2}\,\frac{1}{2}\,\frac{1}{2})$} and {$\mathbf{k}_2=(0\,0.625\,0)$}, with weaker spots in a few other locations, suggesting a complex, potentially multi-$\mathbf{k}$ structure for \mshD\ in its short-range magnetically ordered state at 20\,mK. These wavevectors are consistent with theoretical predictions for the fcc lattice which include third-nearest-neighbor ($J_3$) exchanges~\cite{Balla2020}, implying a key role for these interactions. The resulting magnetic supercell of $13 \times 13 \times 13$ unit cells is available in the 
%Supplemental Material online~\cite{SuppMnSn}.
online ancillary files (see Appendix~\ref{supp}).

The correlation length $\xi$ is typically characterized as the length scale over which the magnetic spins can perceive one another. It can be extracted by considering the absolute value of the spin-pair correlations and modeling them with the Ornstein-Zernike equation \cite{Kim1998},
\begin{align}
\lvert \langle \vec{S}(\mathbf0) \cdot \vec{S}(\mathbf r) \rangle \rvert = {\rm e}^{-r/\xi}.
\end{align}
 Fitting this model to the data in Fig.~\ref{Short_range}(c) yields \mbox{$\xi = 24.66$\,\AA}, corresponding to approximately three unit cells. Additionally, fitting the diffuse magnetic peaks with a Lorentzian profile yields correlation lengths in the range of 10-18\,\AA.

Representative examples of the leading magnetic exchange pathways are schematically illustrated in Fig.~\ref{Magnetic_Pathway}, where Mn atoms are shown connected via the $J_1$, $J_2$, and $J_3$ interactions. These effective couplings arise from indirect super-superexchange mechanisms—often passing through disordered hydrogen bonding or bridging ligands—in the absence of direct corner-sharing Mn–O–Mn linkages. The nearest-neighbor $J_1$ pathway in Fig.\ref{Magnetic_Pathway}(a) is Mn--O--D$\cdots$O--Mn, passing through the disordered hydrogen network. With three intervening atoms and a hydrogen bond, this exchange is likely relatively weak, consistent with the weak nearest-neighbor spin-spin correlations observed in Fig.~\ref{Short_range}(c).  The second-nearest-neighbor exchange $J_2$, illustrated in Fig.\ref{Magnetic_Pathway}(b), takes the form \text{Mn--O--Sn--O--Mn}, and does not pass through protons or hydrogen bonds. This and bond angles closer to 180$^\circ$ are likely to make it more robust than $J_1$ and somewhat less impacted by hydrogen disorder, helping explain the stronger negative spin-spin correlations in Fig.\ref{Short_range}(c), but since both bridging oxygens have disordered protons bonded to them, the disorder will nonetheless have an effect. The third-nearest-neighbor exchange $J_3$ in Fig.\ref{Magnetic_Pathway}(c) is mediated via a more extended Mn--O--Sn--O--D$\cdots$O--Sn--O--Mn pathway. The number of intervening atoms and the involvement of hydrogen bonds via disordered protons are expected to make this exchange much weaker than $J_2$, consistent with the near-zero spin–spin correlations at these distances in Fig.~\ref{Short_range}(c).

\section{Conclusion}

In summary, tetragonal \mshH\ was successfully synthesized, and its crystal structure and magnetic and thermodynamic properties were thoroughly examined through x-ray and neutron diffraction, magnetization, ac susceptibility, and specific heat measurements. Our magnetization, ac-susceptibility and specific-heat results are consistent with the absence of LRO in the compound, while neutron diffraction reveals diffuse magnetic scattering, indicating short-range correlations with a correlation length $\xi = 24.66$\,\AA\ at 20\,mK and magnetic propagation vectors \mbox{$\mathbf{k}_1=(\frac{1}{2}\,\frac{1}{2}\,\frac{1}{2})$} and {$\mathbf{k}_2=(0\,0.625\,0)$}. The magnetic correlation length is significantly shorter than that of the crystal structure, excluding the effect of finite-sized grains or twins.  It remains an open question to what extent the correlated hydrogen disorder is responsible for this. These findings pave the way for further experimental investigations, particularly with local probes such as muons or nuclear magnetic resonance, to explore the static and dynamic ground state of \mshH. Pressure-dependent studies could help identify proton ordering in this disordered hydrogen bond network. Additionally, while we can argue which of the first three exchanges are likely to dominate or be more susceptible to hydrogen disorder, detailed theoretical studies are needed to  model the actual underlying exchange interactions and elucidate the origin of the disordered ground state.
  
\section*{Data Availability}

Samples and data are available upon reasonable request from D.~C. Peets or D.~S. Inosov; data underpinning this work is available from Ref.~\onlinecite{datasetMn}.
%NPD data collected at ANSTO will be made publicly available after a three-year embargo period, i.e., in December 2026, if they ever secure the funding to actually implement this.

\begin{acknowledgments}
We gratefully acknowledge E.\ Hieckmann for her valuable assistance with the SEM/EDX measurements, and L.\ Zviagina and Y.\ Skourski for their help with the pulsed-field magnetization experiments. This project was funded by the Deutsche Forschungsgemeinschaft (DFG, German Research Foundation) through: individual grants IN 209/12-1, DO 590/11-1 (Project No.\ 536621965), and PE~3318/2-1 (Project No.\ 452541981); through projects B03, C01, C03, and C06 of the Collaborative Research Center SFB~1143 (Project No.\ 247310070); and through the W\"urzburg-Dresden Cluster of Excellence on Complexity and Topology in Quantum Materials\,---\,\textit{ct.qmat} (EXC~2147, Project No.\ 390858490). The PPMS at TUBAF was funded through DFG Project No.\ 422219907. The authors acknowledge the support of the Australian Centre for Neutron Scattering (ACNS), ANSTO and the Australian Government through the National Collaborative Research Infrastructure Strategy, in supporting the neutron research infrastructure used in this work via ACNS proposal 16532. We acknowledge support of the HLD at HZDR, a member of the European Magnetic Field Laboratory (EMFL). 
\end{acknowledgments}

\appendix
\section{Crystal Structure Refinement Details\label{appA}}

\begin{table}[bt]
  \caption{\label{Summary}Summary of crystal structure refinements of \mshD\ from powder x-ray and neutron diffraction at 300\,K.}
  \begin{ruledtabular}
    \begin{tabularx}{\columnwidth}
    {l@{\extracolsep{\fill}}l@{\extracolsep{\fill}}l}
      & X-ray & Neutron \\ \hline
      Space group & $P4_2/n$ (\#\ 86) & $P4_2/n$ (\#~86)\\
      $a$ (\AA) & 7.8744(1) & 7.8758(4)\\
      $c$ (\AA) & 7.8242(2) & 7.8247(9)\\
      $V$ (\AA$^3$) & 485.149(3) & 485.352(3)\\
      $Z$ & 4 & 4\\
      Density (g\,cm$^{-3}$) & 3.692(4) & 3.847(1)\\
      $2\theta$ range ($^\circ$) & 4.0--73.3 & 4.0--163.9  \\
      $R$ & 7.54\,\% & 1.91\,\%\\
      $wR$ & 9.68\,\% & 2.44\,\%\\ 
    \end{tabularx}
  \end{ruledtabular}
\end{table}

\begin{table}[tb]
  \caption{\label{XRD_Stoe}Refined atomic positions in \mshD\ from powder x-ray diffraction using 0.559-\AA\ x-rays at 300\,K. Sn and Mn are at $4c$ and $4d$ Wyckoff positions, respectively, while all other atoms are at $8g$.}
\begin{tabular}{lr@{.}lr@{.}lr@{.}lcc}\hline\hline
    Site & \multicolumn{2}{c}{$x$} & \multicolumn{2}{c}{$y$} & \multicolumn{2}{c}{$z$} & $U_\text{iso}$ & Occ. \\ \hline
    Sn1 & 0&5 & 0&0 & 0&5 & 0.004 & 1.0 \\
    Mn1 & 0&5 & 0&0 & 0&0 & 0.004 & 1.0 \\
    O1 & 0&7455(5) & $-$0&0552(2) & 0&5841(1) & 0.004 & 1.0 \\
    O2 & 0&4488(1) & $-$0&2335(1) & 0&6059(5) & 0.004 & 1.0 \\
    O3 & 0&4245(3) & 0&0742(3) & 0&7548(1) & 0.004 & 1.0 \\ \hline\hline
\end{tabular}
\end{table}

\begin{table}[t]
  \caption{\label{NPD_Echidna}Refined atomic positions in \mshD\ at 300\,K from neutron powder diffraction on Echidna using 1.30-\AA\ neutrons. Sn and Mn occupy $4c$ and $4d$ Wyckoff positions, respectively, while all other atoms are at $8g$. The deuteration level refined to 90.38(4)\,\%.}
\begin{tabular}{lr@{.}lr@{.}lr@{.}lcc}\hline\hline
    Site & \multicolumn{2}{c}{$x$} & \multicolumn{2}{c}{$y$} & \multicolumn{2}{c}{$z$} & $U_\text{iso}$ & Occ. \\ \hline
    Sn1 & 0&5 & 0&0 & 0&5 & 0.008 & 1.0 \\
    Mn1 & 0&5 & 0&0 & 0&0 & 0.010 & 1.0 \\
    O1 & 0&7515(5) & $-$0&0637(2) & 0&5894(1) & 0.013 & 1.0 \\
    O2 & 0&4530(1) & $-$0&2348(2) & 0&5813(1) & 0.014 & 1.0 \\
    O3 & 0&4271(5) & 0&0809(2) & 0&7413(1) & 0.011 & 1.0 \\
    D1 & 0&7412(2) & $-$0&1958(1) & 0&5732(6) & 0.033 & 0.5 \\
    D2 & 0&7509(4) & 0&0684(4) & 0&5772(5) & 0.033 & 0.5 \\
    D3 & 0&3279(3) & $-$0&2251(3) & 0&5709(3) & 0.033 & 0.5 \\
    D4 & 0&4383(3) & $-$0&2308(4) & 0&7124(1) & 0.033 & 0.5 \\
    D5 & 0&4224(2) & 0&2023(2) & 0&7447(1) & 0.033 & 1.0 \\ \hline\hline
\end{tabular}
\end{table}

Details of our crystal structure refinements are summarized in Table~\ref{Summary}.  Tables~\ref{XRD_Stoe} and \ref{NPD_Echidna} report the refined atomic positions in \mshD\ at room temperature based on our data collected using x-rays and neutrons, respectively. 
CIF files describing these refinements are provided in the ancillary files as part of this arXiv submission, see Appendix~\ref{supp}.
%CIF files describing these refinements are provided in the Supplemental Material online\,\cite{SuppMnSn}.

\section{Supplemental Material\label{supp}}

As ancillary files to this arXiv submission, we provide the following crystallographic information files (CIFs) describing our crystal structure refinements:

\begin{center}
\noindent\begin{tabular}{l}
\verb+Stoe_300K_0p559A_MnSn(OD)6.cif+\\
\verb+Echidna_300K_1p3A_MnSn(OD)6.cif+\\
\verb+Echidna_0p020K_2p44A_MnSn(OD)6.mcif+\\
\end{tabular}
\end{center}

\bibliography{MnSnOH6}

\end{document}